\documentclass[aps,superscriptaddress,amsmath,amssymb,twocolumn,amsfonts]{revtex4-1}
\usepackage{amsmath}
\usepackage{graphicx}
\usepackage{hyperref}
\usepackage{cleveref}
\usepackage{color}
\usepackage{float}
\usepackage{soul}
\usepackage{ulem}
\usepackage{braket}

\newcommand\bea{\begin{eqnarray}}
\newcommand\eea{\end{eqnarray}}
\newcommand\beq{\begin{equation}}  
\newcommand\eeq{\end{equation}}

 \usepackage{physics}
\usepackage{tensor,upgreek,float}
 \usepackage{parskip}
\newcommand\Perms[2]{\tensor[^{#2}]C{_{#1}}}
 
\begin{document}

\title{Charge Transfer Energy \& Band Filling Effects on Two-Hole Auger Resonances in strongly correlated systems.}
\author{Prabhakar}
\author{Anamitra Mukherjee} 
\affiliation{School of Physical Sciences, National Institute of Science Education and Research, a CI of Homi Bhabha National Institute, Jatni 752050, India}
\date{\today}
\begin{abstract}
We investigate the impact of charge transfer energy and band-filling on the stability of the two-hole resonance relevant for Auger electron spectroscopy (AES) in transition metal oxides. 
As a minimal model to study charge transfer effects in a transition metal (TM) and Oxygen (OX) chain, we consider a one-dimensional chain with spinless fermion with an alternating motif of site-pairs with nearest neighbor (NN) repulsion $U$ and uncorrelated site-pairs,  separated by a charge transfer gap $\Delta$. \textcolor{black}{We first show that while two holes added in a filled band of  NN interacting fermion in one dimension can stabilize to a two-hole bound pair, the bound pair delocalizes with a $U$-dependent bandwidth.} In contrast, we establish that the bandwidth of two holes added on a TM site-pair in a filled band is dramatically suppressed,  realizing a `local' two-hole resonance (L2HR) at the same TM site-pair mimicking the AES phenomenology. Employing a memory-efficient exact numerical scheme and standard Lanczos-based diagonalization, we then study two-hole spectra for holes added at TM site-pairs in partially filled bands. We analyze the multiple features that arise in the two-hole spectra at partial filling of the ground state. We uncover that in the strong $U$ limit, there is a filling-dependent $\Delta_{crit}$ above which the L2HR remains stable for any band-filling greater than 75\%. In this regime, the energy location of the L2HR provides a direct estimate of the correlation strength at TM site-pairs for partial filling and is reminiscent of the Cini-Sawatzky theory for the filled band case. At 75\% band-filling, an abrupt redistribution of two-hole spectral weight destroys the L2HR regardless of  $U$ or $\Delta$ values. 
We discuss the relevance of these nonperturbative results, obtained with full lattice symmetry, for understanding AES of partially filled bands in terms of the local-two-hole spectrum. 
\end{abstract}

\maketitle

\section{Introduction}
Few-body bound states are of fundamental importance in the study of many-body physics. They lead to emergent properties that include superconductivity mediated by Cooper pairs, repulsively bound pairs in Bose-Hubbard models \cite{bhm-zoller}, trion bound states in GaAs quantum wells \cite{gaas}, Efimov states \cite{efimov,exp-efimov}, and stable molecules. When these bound complexes are exact eigenstates, they represent true bound states. However, in many situations, the low energy states are described by bound complexes interacting with a continuum of excitations, where the complexes acquire a finite lifetime and occur as resonances. Such resonances also occur in the study of two-hole bound complexes in the Auger electron spectroscopy (AES) of correlated materials\cite{RevModPhys.93.035001,aug-prl-1,aug-prl-2,aug-nat,auger-old-2}. 


The `Core Valence Valence' or CVV Auger process consists of the decay of an X-ray-induced core hole into two final state valence holes plus an Auger electron and is mediated by on-site Coulomb interactions \cite{auger-old-1, aes-cini, auger-old-2}. 
AES provides crucial insights into local atomic multiplet structure, on-site interaction strengths, and crystal fields\cite{three, four, five}. Further, supplementing AES of the transition element with Oxygen KLL Auger spectra yields additional information on Oxygen on-site repulsion energy and interactions between holes in neighboring transition-metal and Oxygen ions\cite{six}. Such information is vital for understanding correlated materials, making AES a longstanding focus of extensive research \cite{PhysRevLett.121.233201, PhysRevResearch.3.033063, PhysRevLett.107.233001, PhysRevA.79.022508, PhysRevA.77.053404, PhysRevLett.71.4307, PhysRevLett.72.621, PhysRevLett.74.2917, PhysRevLett.76.3100}. 

\textcolor{black}{The  well-known exact solution to the two-hole Green's function, the Cini-Sawatzky\cite{aes-cini,auger-old-1} theory, applies to simple cases, e.g., Cu, Zn where two holes are added in a full $3d$ band.} In this case, the central quantity of interest, the local two-hole Green's function, can be computed from non interacting single-hole Green's function. The resulting local two-hole spectral function weighted with suitable matrix elements provides excellent agreement with experiments. Further generalizations to include dynamical screening, off-site interactions, overlap effects, and one-step formulation have helped sharpen the theoretical analysis \cite{auger-old-2,one-step}.  
However, in partially filled bands, success has been limited. 
The Green's function of two holes added at an atomic site in the background of holes in the partially filled ground state is a non-trivial many-body problem. Early attempts include ladder approximation\cite{lda-1,lda-2,3-ladder} and diagrammatic vertex correction\cite{POTTHOFF1995163} approaches. They have achieved some success for partial filling close to the fully-filled limit, by adding small number of holes in the filled band. Another standard approach to the problem is the impurity approximation. Here a transition metal oxide with transition metal (TM) ion intercalated with Oxygen (OX) is simplified to an Anderson impurity problem of a single TM ion connected to an OX lattice\cite{dd-auger, PhysRevB.44.9656, PhysRevB.33.8060, PhysRevB.28.4315}. More recently, variational approaches for computing few-body bound states in one dimension have been developed\cite{cfgf-anam, cfgf-all}. 
These approaches have added valuable insights to the two-hole propagator in partially filled bands. However, these methods have shortcomings that need to be alleviated. Impurity models are computationally less intensive and can capture local atomic physics but lower the true lattice symmetries and do not capture the effect of band filling exactly. The many-body perturbative approaches are restricted to partial filling close to filled ground state. Finally, the variational approach developed for few-body Green's function is difficult to generalize to finite hole-doped ground states. 

Here, we present numerically exact results of two-hole spectral functions for interacting spinless fermions in one dimension. These results go beyond perturbative approaches, retain full lattice symmetry unlike impurity approximations, and can be employed at any partially filled ground state. \textcolor{black}{A repulsively two-hole bound state can be stabilized for spinless fermions with  nearest neighbour (NN) repulsion. Such a bound state delocalizes with a correlation-dependent bandwidth. The reason for considering spinless fermions is to limit the Hilbert space dimension to access larger system sizes. Since a single spinless fermion can occupy each site, the two-hole bound state comprises two holes on NN sites instead of being localized on a single site in actual AES. However, as discussed in the paper, the stability of the two-hole bound state is largely unaffected by the inclusion of spin degrees of freedom. \\
Another important experimental fact is that, the two-hole bound state in AES remains localized near the atom where the core hole is created over experimental time scales. 
We consider a one-dimensional model with pairs of TM sites with NN interaction $U$ separated by pairs of OX sites and add a charge transfer energy $\Delta$ between the TM and OX site-pairs. We show that it is possible to drastically suppress the bandwidth of the two-hole bound state within our model, approximating almost localized two-hole bound state. This `local' two-hole resonance within our model implies a pair of holes localized on a TM site-pair. The nomenclature of TM and OX is borrowed from transition metal and oxygen sites, respectively, in a transition metal oxide chain. These energy scales are a natural first choice to investigate as, in transition metal oxides, Coulomb interactions and charge transfer energy are dominant energy scales, for example, facilitating the well-known Zaanen-Sawatzky-Allen classification\cite{zsa} and are also known to be the main driver two-hole resonance in AES \cite{aes-cini,auger-old-1}. Restricting to only these scales allows us to uncover how ground state charge fluctuation affects the Green's function of two holes added at a TM site-pair over a wide range of hole-doped ground states. We investigate the model within an exact numerical scheme\cite{prabhakar2022memory} and standard Lanczos-based diagonalization.\\}

We first compute the two-hole spectral function in a filled ($n=1$) lattice of spinless fermions with NN interaction $U$ to set the stage. From this, we establish that two holes added on NN sites delocalize as a pair with a narrow $U$-dependent bandwidth, forming a two-hole bound pair beyond a critical $U$. We then consider a model that mimics the phenomenology of AES by creating a chain of uncorrelated OX  site-pairs and NN interacting  TM site-pairs with a correlation strength $U$ and a charge transfer energy  $\Delta$ between OX and TM. \textcolor{black}{We show that, unlike the simple NN repulsive model for $n=1$, the two-hole bound pair is strongly localized (with vanishingly small bandwidth)  at the TM site-pair where they are initially added.} However, its stability requires both large $U$ and $\Delta$. We refer to this resonance as a `\textit{local}' two-hole resonance (L2HR), indicating two holes localized on a single TM site-pair. We then investigate the interplay of $\Delta$ and band filling ($n$) in the strong interaction regime or for appropriately chosen large $U$ values.

We detail how partial filling $(n<1)$ of the ground state and $\Delta$ play out in the large $U$ limit by mapping out the $\Delta-n$ parameter regime where the L2HR is stable. We establish that the L2HR is stable for $0.75< n \leq 1$, provided $\Delta$  is greater than a filling-dependent  $\Delta_{crit}$. By examining the detailed evolution of the two-hole spectral function with ground state filling and $\Delta$, we provide the reason for the observed limited range of stability in the  $\Delta-n$ parameter space and the sudden redistribution of spectral weight at $n=0.75$ that destroys the signature of the L2HR in the local two-hole spectral function. We show that in the regime where the L2HR is stable at partial band filling, its energy location varies linearly with $U$ and $\Delta$, similar to the Cini-Sawatzky theory for L2HR in filled bands. We conclude by commenting on the relevance of our results for Auger electron spectroscopy.

The paper is organized as follows. We summarize the theoretical approach for analysing AES in section-II. Section III briefly discusses the many-fermion formalism for extracting the two-hole spectral function in real space. In section IV-A, we discuss the results of the NN interacting fermions. We discuss the results of the interplay of charge transfer, interaction, and doping on the stability of the L2HR for the TM and OX site-pair model in section IV-B. We conclude the paper in Section V.

\section{General theory of Auger electron spectroscopy}

\textcolor{black}{We briefly discuss the theory of core-valence-valence (C-VV) Auger scattering. The  Auger scattering is a radiationless process in which an X-ray-induced core hole at an atomic site produces two valence band holes along with the emission of an Auger electron. Within the two-step model\cite{aes-cini,auger-old-1}, the initial ionization of the atom by X-ray and the Auger relaxation are treated independently. 
The initial state consists of a filled valence shell and an atomic site with a core hole. Assuming almost instantaneous thermalization of the core hole with its environment allows approximating the initial state energy $E_{i}$ to be equal to $E^{\mathcal{N}}_0+E_c$. Here,  $E_c$ is the core-hole energy, and $E^{\mathcal{N}}_0$ is the ground state energy of a Hamiltonian $H_v$ describing the valence band electrons including kinetic energy and electron repulsion. In the Auger relaxation, a valence electron fills the initial core-hole state, and the energy released excites another valance band electron in a scattering state with energy $E_{\mathbf{k}}$. The sudden approximation allows one to treat the two-hole valence state and the emitted electron as independent processes. Thus, the final state energy can be identified with $E_{f}=E^{\mathcal{N}-2}+E_{\mathbf{k}}$ where, $E_{\mathcal{N}-2}$, is the $(\mathcal{N}-2)$-electron excited state of the valance band Hamiltonian $H_v$.\\
The Auger transition is determined by  the matrix element \cite{two-step-3} $\int  \varphi_{c\sigma}^*(\mathbf{R_I}-\mathbf{r_1}) \varphi_{\mathbf{k\sigma^\prime}}^*(\mathbf{r_2})|\frac{1}{|\mathbf{r_{1}}-\mathbf{r_2}|}|\varphi_{v\sigma^\prime}(\mathbf{R_J}-\mathbf{r_1}) \varphi_{v\sigma}(\mathbf{R_L}-\mathbf{r_2}) d\mathbf{r_1}d\mathbf{r_2}$. The core-hole wavefunction at the atomic location $\mathbf{R_I}$ is denoted by $\varphi_{c\sigma}(\mathbf{R_I}-\mathbf{r_1})$ and a one-body scattering state for the Auger electron is $\varphi_{\mathbf{k}\sigma}(\mathbf{r_2})$. $\varphi_{v\sigma}(\mathbf{R_J}-\mathbf{r_1})$ and $\varphi_{v\sigma}(\mathbf{R_L}-\mathbf{r_2})$ denote the final state valence state wavefunctions of two holes located at atoms at $\mathbf{R_J}$ and  $\mathbf{R_L}$. Usually, the Auger process is dominated by intra-atomic contribution, allowing the simplification of identifying $\mathbf{R_I=R_J=R_K}$. We label this simplified matrix element by, $M_{R_I}^{\sigma\sigma^\prime}(\mathbf{k})\equiv \frac{e^2}{4\pi\epsilon_0}\int  \varphi_{c\sigma}^*(\mathbf{R_I}-\mathbf{r_1}) \varphi_{\mathbf{k}\sigma^\prime}^*(\mathbf{r_2})|\frac{1}{|\mathbf{r_{1}}-\mathbf{r_2}|}|\varphi_{v\sigma^\prime}(\mathbf{R_I}-\mathbf{r_1}) \varphi_{v\sigma}(\mathbf{R_I}-\mathbf{r_2}) d\mathbf{r_1}d\mathbf{r_2}$. In terms of the Fermi-Golden rule, we have the momentum-resolved Auger intensity proportional to $|M_{R_I}^{\sigma\sigma^\prime}(\mathbf{k})|^2\delta(E^{\mathcal{N}-2}+E_\mathbf{k}-E^{\mathcal{N}}_0-E_c)$. Defining $(E_\mathbf{k}-E_c)\equiv \hbar\omega$, we have  $\mathcal{I}_{R_I}(\omega,\mathbf{k})=\sum_{\sigma,\sigma^\prime}|M_{R_I}^{\sigma\sigma^\prime}(\mathbf{k})|^2\delta(\hbar\omega+E^{\mathcal{N}-2}-E^{\mathcal{N}}_0)$. The $\delta-$function contains information of the $(\mathcal{N}-2)$-fermion excitations in $\mathcal{N}$-fermion ground state of $H_v$, generated by two-electrons removed from the site $\mathbf{R_I}$. This local two-hole spectral function can be obtained from the Fourier transform of the corresponding local two-hole Green's function with two holes created simultaneously in the $\mathcal{N}$-fermion ground state of the atom at $\mathbf{R_I}$ and subsequently removed together. Thus, from a many-body perspective, the central quantity of interest is the two-hole local propagator with two-time labels. \\
For a one-orbital lattice Hamiltonian describing a valence band with spin-full fermions and onsite repulsion (Hubbard model), the Green's function to be computed would involve two holes of opposite spins added at the site $\mathbf{R_I}$. In the present study, as discussed in the introduction, we consider spinless fermions with nearest-neighbor interactions in a one-orbital lattice model. Here, the double occupation of a single site is Pauli blocked. Thus, in our model with interacting spinless fermions \textit{two-hole localization refers to holes localized on adjacent lattice sites}. Thus, the relevant Green's function to be computed involves two holes on adjacent sites, as elaborated on in the next section.}


\section{Method}
We consider a periodic lattice of $\mathcal{L}$ sites with $\mathcal{N}$ fermions. Based on the discussion in the previous section, our quantity of interest is the retarded two-hole propagator $G_{IJ;IJ}^{(2h)}(t,t')\equiv\langle\psi_0^\mathcal{N}|c^{\dagger}_{I}(t')c^{\dagger}_{J}(t')c_{I}(t)c_{J}(t)|\psi_0^\mathcal{N}\rangle$, where two fermions are destroyed at sites $I$ and $J$  at a time $t$,  and added back at the same site-pair at a later time $t'$, in the many fermion ground state $|\psi_0^\mathcal{N}\rangle$ with ground state energy $E_0^\mathcal{N}$. The real-space two-hole spectral function in the frequency space $A^{(2h)}_{IJ;IJ}(\omega)$ is given by $-1/\pi Im\{G_{IJ;IJ}^{(2h)}(\omega)\}$, where $G_{IJ;IJ}^{(2h)}(\omega)$ is the Fourier transform of $G_{IJ;IJ}^{(2h)}(t,t')$.  

For an interacting Hamiltonian $H$ defined on a $\mathcal{L}$ site periodic chain with $\mathcal{N}$ fermions, we consider a generic eigenvalue problem $\hat{H}|\lambda^\mathcal{N}\rangle=E^{\mathcal{N}}_{\lambda}|\lambda^{\mathcal
{N}}\rangle$, with $\{E^{\mathcal{N}}_\lambda\}$ and $\{|\lambda^\mathcal{N}\rangle\}$ are the $\mathcal{N}$-particle eigenvalues and eigenvectors, respectively. 
\textcolor{black}{
We have employed atomic units that amounts to setting $\hbar$, electron charge ($e$) , electron mass $m_e$ and $1/4\pi\epsilon_0$ equal to 1. } 
We choose a real space $\mathcal{N}$ fermion basis $\{|j_\mathcal{N}\rangle\}\equiv \{c^{\dagger}_{a_1}..
.c^{\dagger}_{a_\mathcal{N}}|0\rangle\}$, whose elements are generated by permuting real space fermion positions, here denoted by the $a_i$ subscripts of the creation operators. 

\textcolor{black}{We emphasize that the real space basis are constructed from the fermion occupations of lattice sites in real space more appropriately called as site-basis.} 
In the compact $\{|j_\mathcal{N}\rangle\}$ basis notation the $\mathcal{N}$-fermion Green's function is denoted by $ \mathcal{G}^{\mathcal{N}}_{i_{\mathcal{N}};j_{\mathcal{N}}}(\omega)$. We note that the $\mathcal{N}$-fermion Green's function is a matrix $[\mathcal{G}^{\mathcal{N}}(\omega)]$ in the $\{|j_\mathcal{N}\rangle\}$ basis and defines the $\mathcal{N}$-fermion spectral function matrix by the relation $[\mathcal{D}^{\mathcal{N}}(\omega)]\equiv -1/\pi \{Im[\mathcal{G}^{\mathcal{N}}(\omega)]\}$. A trace of the $\mathcal{N}$-fermion spectral function matrix $[\mathcal{D}^{\mathcal{N}}(\omega)]$ taken over the $\mathcal{N}$-fermion basis provides the $\mathcal{N}$-fermion density of states $\mathcal{A^{N}}(\omega)$. 

For two holes created at site-pair $(I,J)$ in the many fermion ground state $|\psi_{0}^{\mathcal{N}} \rangle$ and subsequently destroyed from the same locations, the real-space two-hole spectral function is given by: 
\begin{eqnarray}
A^{(2h)}_{IJ;IJ}(\omega)&=&\sum_{j_{_{\mathcal{N}}},j^\prime_{_{\mathcal{N}}}
}\mathcal{D}^{\mathcal{N}}_{j_{_{\mathcal{N}}},j^\prime_{_{\mathcal{N}}}}(E^
{\mathcal{N}}_0)\mathcal{D}^{\mathcal{N}-2}_{j_{_{\mathcal{N}}}(IJ)^-;j
^\prime_{_{\mathcal{N}}}(IJ)^-}(\omega)~~~~~\newline
\label{e3}
\end{eqnarray}
The details of the derivation are standard and are provided in Appendix-1. From the formula, we see that the two-hole spectral function is expressed in terms of elements of the $\mathcal{N}$ and $\mathcal{N}-2$  fermion spectral function, $[\mathcal{D}^{\mathcal{N}}(\omega)]$ and $[\mathcal{D}^{\mathcal{N}-2}(\omega)]$ respectively. The relation $|j_\mathcal{N}(IJ)^-\rangle=c_Ic_J|j_\mathcal{N}\rangle$ defines the indices of $[\mathcal{D}^{\mathcal{N}-2}(\omega)]$ in Eq.~\ref{e3}. The definitions for the primed labels are analogous. The elements of $[\mathcal{D}^{\mathcal{N}}(\omega)]$ are needed at $\omega=E^{\mathcal{N}}_0$, the $\mathcal{N}$-fermion ground state energy. $E^{\mathcal{N}}_0$ is determined from the lowest energy peak of the $\mathcal{N}$-fermion density of states $\mathcal{A^{N}}(\omega)$. 
Different elements of $\mathcal{D}^{\mathcal{N}}_{j_{_{\mathcal{N}}},j^\prime_{_{\mathcal{N}}}}(E^{\mathcal{N}}_0)$ are extracted from the $\mathcal{N}$-fermion Green's function $\mathcal{G}^{\mathcal{N}}_{j_{_{\mathcal{N}}};j^\prime_{_{\mathcal{N}}}}(\omega)=\langle j_{_{\mathcal{N}}}|\mathcal{\hat{G}(\omega)} |j^\prime_{_{\mathcal{N}}}\rangle$, evaluated between the $\mathcal{N}$ particle basis elements, $(|j_{_{\mathcal{N}}}\rangle)^\dagger$ and $|j_{_{\mathcal{N}}}^\prime\rangle$. We calculate $[\mathcal{G}^{\mathcal{N}}(\omega)],[\mathcal{G}^{\mathcal{N}\pm 1}(\omega)]$ and $[\mathcal{G}^{\mathcal{N}-2}(\omega)]$ using a recently developed memory-efficient variant of full exact diagonalization. The scheme is outlined in Appendix II; we refer the reader to our recent work for details and numerical benchmarks \cite{prabhakar2022memory}. 

\begin{figure*}[t]
\centering{
\includegraphics[width=18cm, height=5.7cm, clip=true]{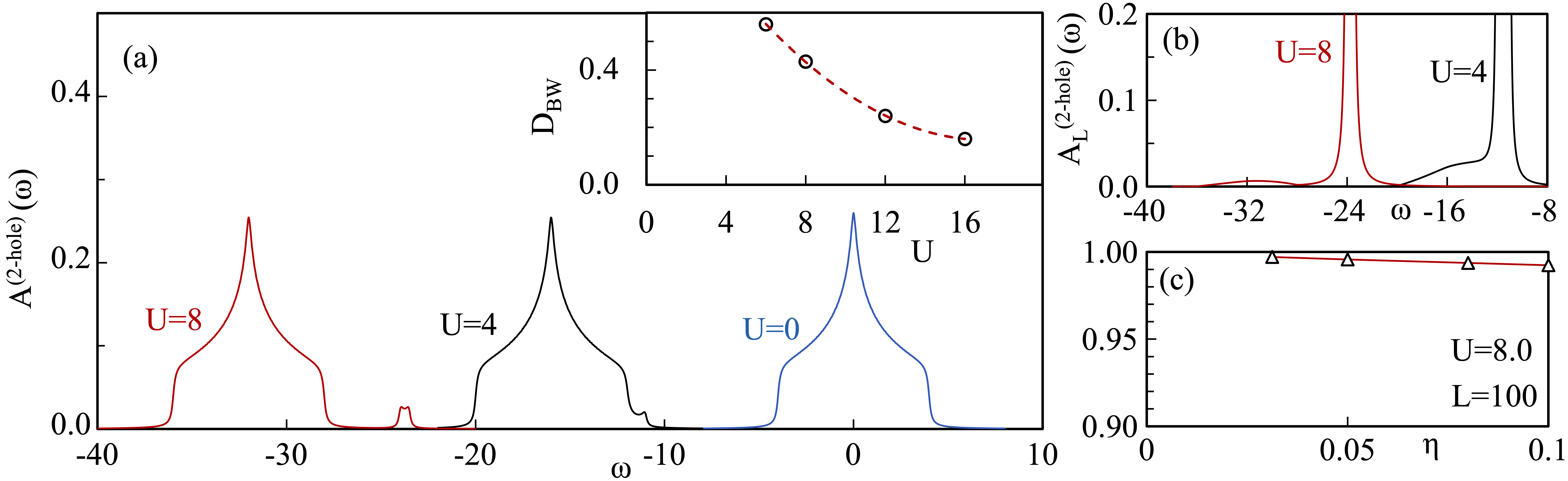}}
\caption{\textbf{Two hole excitation spectrum in a filled ground state for spinless fermions with NN interacting:} The main panel (a) shows the two hole spectral function, $A^{(2-hole)}(\omega)$ for indicated $U$ values. The dissociation of the single continua into two distinct features is shown for $U=8$. \textcolor{black}{Inset in (a) shows the comparison of the numerical bandwidth of the two-holes delocalizing as a bound pair  (circles) and analytic scaling (dashed line). }(b) shows the local two-hole spectral function $A_L^{(2-hole)}(\omega)$ for $U=4$ and 8. (c) shows the ratio of the numerical and exact frequency two-hole spectral sum rule value as a function of broadening factor $\eta$. All results are for $\mathcal{L}=100$ site chain with periodic boundary conditions.}
\label{f-2}
\end{figure*}

\section{Results}

\subsection{Repulsive spinless-fermions in 1D}
We first study the spectral response of a filled $(\mathcal{L}=\mathcal{N})$ or $n=1$ many-body ground state to the introduction (and subsequent removal) of two holes in the spinless-fermion model with NN repulsion. We define the model with periodic boundary condition (pbc) on a $\mathcal{L}$ site chain with NN interactions, as follows:
\begin{equation}
 H=-t\sum\limits_{\langle I,J\rangle }(c_{I}^{\dagger}
c_{J}+h.c)+U\sum\limits_{I} n_{I} n_{I+1}
\label{e7}	
\end{equation}
where $c^\dagger_{I}$ ($c_{I}$) are spinless fermion creation (annihilation) operators at the site $I$. $t$ and $U$ are the NN  hopping and interaction, respectively. $n_{I}=c^\dagger_{I}c_{I}$ is the number operator at a site $I$. We measure all energies in units of $t$, which we have set to unity. 
We analyze the properties of two holes created in a filled system ground state by calculating the \textit{total} two-hole spectral function, $\sum_{I\neq J}A^{(2-hole)}_{IJ;IJ}(\omega)\equiv A^{(2-hole)}(\omega)$. We also calculate the \textit{local} two-holes spectral function $A^{(2-hole)}_{IJ;IJ}(\omega)\equiv A_L^{(2-hole)}(\omega)$ for holes created on NN sites I and J. In the present case, the term `local' implies the spectral function for holes on NN sites, the closest analog of onsite in the spin-full problem.

Fig.~\ref{f-2}(a) shows $A^{(2-hole)}(\omega)$ for different values of $U$ where the spectrum is shifted by the ground state energy $E_0^\mathcal{N}(=\mathcal{N}U)$. For $U=0$, we have the two-hole continua extending from $\omega\approx -4t$ to $\omega\approx4t$. This feature is the two-hole band comprising of the holes moving independently respecting the Pauli exclusion principle and agrees with the analytical form $\sum_{k\neq k^\prime}\delta(\omega+2tcos(k)+2tcos(k^\prime))$, with the $k$ ($k^\prime$) sums extending over the entire Brillouin zone $[-\pi,\pi)$. For $U=4$, we see that the band center is at $\omega\approx -16$ or $4U$ below zero. We also see that the two-hole spectral function is distorted compared to the $U=0$ case, and a small feature appears at the upper band edge. Beyond $U=5$, this feature splits off, creating a band with two holes delocalizing as a pair. We show typical data for $U=8$, where we find a band of width $8$, centered around $\omega=-32$, and a narrow band centered at $\omega=-24$. To understand the spectrum, we analyze the \textit{potential energy of the basis states}. The potential energy is the correlation energy of the fermion configurations in a basis state in the limit of no hopping. For all basis states containing the two holes on nearest neighbor sites, the potential energy is $(\mathcal{N}-3)U$, while for all other two-hole basis states, it is $(\mathcal{N}-4)U$. Since we shifted the spectra by the ground state energy ($\mathcal{N}U$), the centroids of the two bands with states containing two holes not on nearest neighbor sites are at -$4U$. Similarly, the states with two holes on NN sites contribute to the band centered around -$3U$. The finite bandwidths result from kinetic energy that gives a width of $8t$ for the two holes delocalizing independently and a much smaller bandwidth to the other band. Inset in panel (a) shows the scaling of the bandwidth of the narrow band ($D_{BW}$) as a function of $U$ by open symbols. The solid lines are results from an analytical projection approach that agrees well with the numerical results. Details of the analytical calculation are presented in Appendix-3.

To further clarify the character of the split off-peak, in panel (b), we show $A^{(2-hole)}_{L}(\omega)$, for holes created (and subsequently destroyed) on a pair of adjacent sites $I$ and $J$. For $U=4$, we find a wide distribution of spectral weight over the entire width of the total two-hole spectral function. Significant overlap exists between the sharp peak near $-10$ and the broad continua. 
In contrast, for $U=8$, we see a clear split-off feature at $U=-24$ and well-separated, band-like continua coinciding with the two-hole band centered around $-32$ in (a). The split-off feature has the same width as $D_{BW}$ (Inset in (a)), showing that the two holes delocalize as a pair. \textcolor{black}{However, it has a small overlap with the broad band centered at $-32$, implying that this feature  is a signature of a two-hole\textit{ bound pair} has a finite lifetime. }
\textcolor{black}{Similar repulsively bound states have been reported for bosons with $U>>t$ \cite{bhm-zoller, pothoff-2} and the spin-full Hubbard model within DMRG \cite{dmrg}.}

\textcolor{black}{We would like to briefly discuss the numerical accuracy of these results. As is well known that from analytic properties of Green's function, shifting the poles of the Green's function above and below by a regulator $\eta$ the real axis defines the retarded  and advanced Green's function. The expression for the retarded Green's function with a finite $\eta$ is provided in Eq. (A.1). Thus, our two-hole spectral function derived from the Green's function depend on $\eta$. A standard way to  understand the systematics of $\eta$-dependence of the two-hole spectral function, is to compute the two-hole frequency sum rule for different values of $\eta$ and compare the result with the theoretical value of  $\mathcal{L}(\mathcal{L}-1)$. }
Hence in (c), we show the two-hole frequency sum rule plot as a function of $\eta$. It is defined as $\int_{-\infty}^{\infty}A^{(2-hole)}(\omega)d\omega$, whose exact analytical value is   $\mathcal{L}(\mathcal{L}-1)$, for two holes in a filled band. The ratio of the numerical value to $\mathcal{L}(\mathcal{L}-1)$ rapidly approaches 1,  as $\eta\rightarrow 0$.

\textcolor{black}{Before we move on to the main model  investigated in our paper,  we comment on the essential difference between the spin-full fermion and the present study with spinless fermions. We  emphasize that the main question being addressed in this paper is the validity of the local two-hole spectral function as a signature of Auger spectroscopy in partially hole-doped bands. Spin physics is not expected to play a dominant role in determining the critical parameter values of the location of the split-off feature. For example the  location of the split-off feature in the main panel of  Fig. 2 (b)  is given by $-2\Delta-U$. This agrees closely with previous work with two-body spin-full fermions split-off feature in \cite{cfgf-anam}. }
 
\subsection{Interplay of band-filling \& charge transfer energy}
We now consider the impact of band-filling and charge transfer effects in the presence of strong interactions on the local two-hole spectral function. The model is defined as follows:
\begin{eqnarray}
H&=&-t\sum\limits_{\langle I,J\rangle }(c_{I}^{\dagger}
c_{J}+h.c)+U\sum\limits_{I=1}^{\mathcal{L}/4} n^{TM}_{4I-3} n^{TM}_{4I-2}\nonumber\\ &+&\Delta\sum\limits_{I=1}^{\mathcal{L}/4} (n^{TM}_{4I-3}+n^{TM}_{4I-2})
\label{e-h2}
\end{eqnarray}

\begin{figure}[t]
\centering{
\includegraphics[width=8.5cm, height=8.5cm, clip=true]{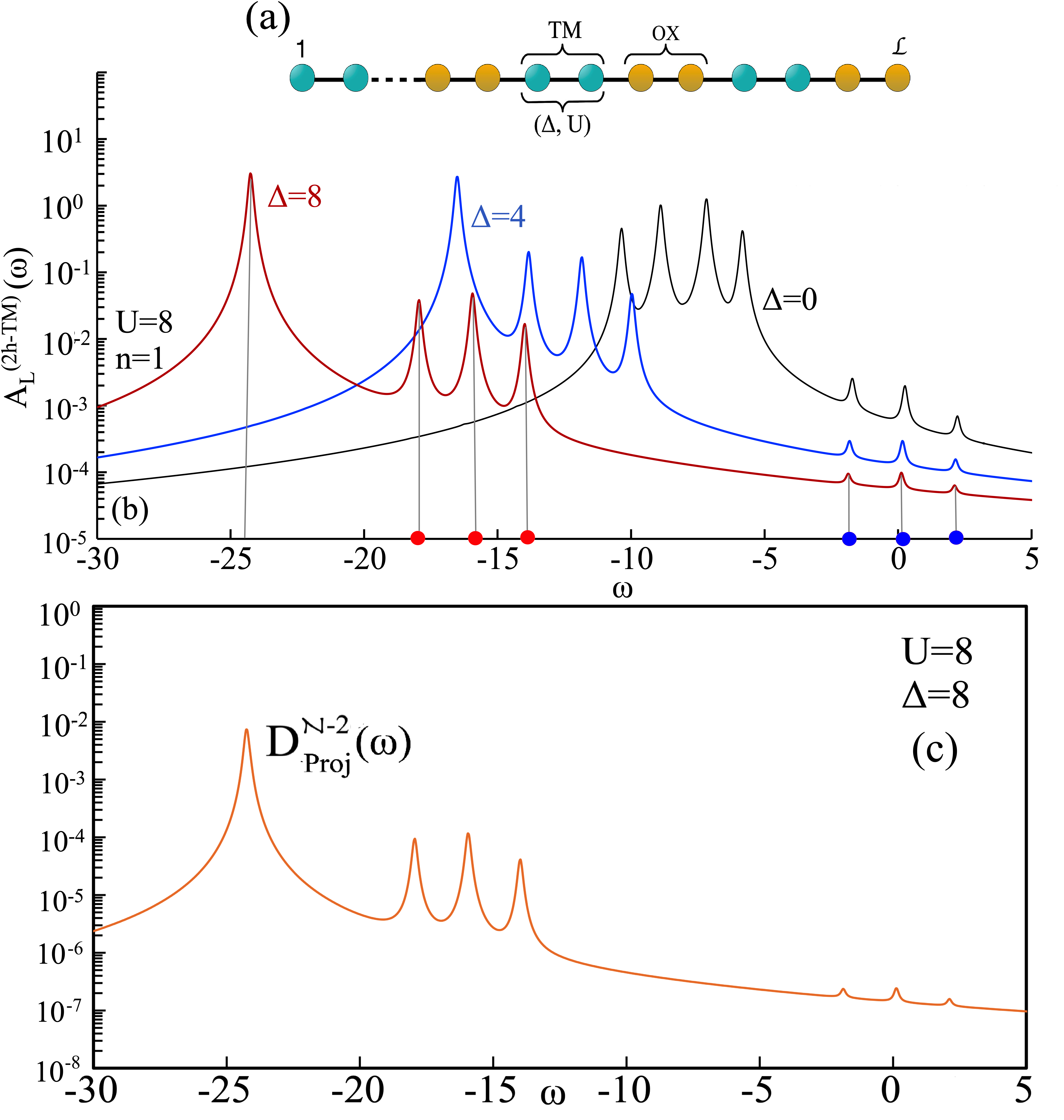}} 
\caption{\textbf{\textit{Local} two-hole spectral function at TM site-pair in filled bands.} (a) Schematic of the periodic model. (b) the local two-hole spectral function for pair of holes created on a TM site-pair for $n=1$ and three $\Delta$ values as indicated. Results shown are for $U=8$ and $\mathcal{L}=100$. The vertical lines for $\Delta=8$ mark the locations of the centers of the three features. The red and blue ellipses on the $\omega$-axis are discussed in the text. (c) shows projected $\mathcal{D}^{\mathcal{N}-2}_{Proj}(\omega)$ which quantifies the contribution to the $(\mathcal{N}-2)$-fermion spectral function from basis states with two holes doubly occupying any TM sites-pair.}
\vspace{-0.0cm}
\label{f-3}
\end{figure}

In Fig.~\ref{f-3} (a), we show a schematic of the Hamiltonian. \textcolor{black}{The chain contains pairs of sites labelled as TM sites with onsite energy $(\Delta\geq 0)$ and  nearest neighbor (NN) repulsion strength of $U$ among them. The TM site- pairs are separated by pairs of  sites labelled as OX sites with $U=0$. $\Delta$ acts as a charge transfer energy between the TM and OX sites. Thus the TM site-pair with NN interactions acts as the simplest extension of alternating chain of TM and OX with spin-full fermions . The spinless-fermion model allows us examine the formation of spatially-localized two holes on adjacent lattice sites as opposed to onsite localization in the spin-full fermion case. Below we show that indeed such two-hole resonances are stabilized in certain situations. We refer to such two holes resonances as `local' two-hole resonance (L2HR). As pointed out  at the end of the previous section, the locations of the two-hole spectral function features remain largely unaffected by spin excitations. Further,  inclusion of spins severely restricts the lattice sizes that can be accessed due to significantly enlarged Hilbert space. Due to these reasons we choose to work with the spinless fermion model.}


\textcolor{black}{The creation (annihilation) operators have the usual meaning. $I,J$ in the kinetic energy term runs over  NN sites. $n^{TM}_I$ is the number of operators on the TM site $I$. The OX sites are non-interacting sites. We also  define $n^{OX}_I$ as the number of operators on the OX site $I$. With these identifications we define  $n=\frac{1}{\mathcal{L}}(\sum_{I}n_I^{TM}+\sum_{I}n_I^{OX})$, where the sum runs over TM and OX sites in the first and second summations respectively. Here $\mathcal{L}$ refers to the total number of sites. In terms of these we define fully filled ($n=1$) and partially filled bands ($n<1$). This $n=1$ refers to a system where all TM and OX sites are fully occupied. Finally the hopping between all sites is $t$. }We study local two-hole spectra for two holes created on adjacent TM sites in the filled ($n=1$) and in partially filled ($n<1$) or $(1-n)$ hole-doped ground states on $\mathcal{L}$ sites with pbc.  For the present model, the `local' two-hole spectral function of interest is for two holes created on a single TM site-pair, as mentioned above. \textcolor{black}{These TM-holes hybridize with holes states in OX  sites and can delocalize through the system.}

\underline{\textit{(i) Local two-hole spectrum in the undoped case:} }
In Fig.~\ref{f-3} (b), we show the local two-hole spectral function defined as $A^{(2-hole)}_{IJ;IJ}(\omega)\equiv A_{L}^{(2h-TM)}(\omega)$, where $I$ and $J$ are adjacent sites belonging to a TM site-pair. The result is shown for $U=8$ and three values of $\Delta$ for a 100-site lattice. The results are presented in a logarithmic scale on the y-axis to show the various structures clearly. The ground state energy for the fully filled case is $E_0^\mathcal{N}=\mathcal{L}(U+2\Delta)/4$ as $U$ and $\Delta$ are present only for the TM sites. As above, all the results are shifted by $E_0^\mathcal{N}$.  
For $\Delta=8$, we see a well-separated feature at $\omega=-2\Delta-U(=-24)$. This energy corresponds to the potential energy reduction at $t=0$ measured from the ground state potential energy. It is identical for all basis states, with the two holes doubly occupying a TM site-pair. The next structure is centered around $\omega=-\Delta-U(=-16)$, the potential energy of basis states with one hole on the TM site and one in the OX sites. Finally, the highest energy feature centered around $\omega=0$ refers to the case where both holes are on OX sites. We notice that the second and third features, respectively, are two and five orders of magnitude smaller than the first feature. To go beyond the simple potential energy-based analysis, we note that for the filled case $\mathcal{D}^{\mathcal{N}}_{j_{_{\mathcal{N}}},j^\prime_{_{\mathcal{N}}}}(E^
{\mathcal{N}}_0)=1$ as there is only one state, the filled many body configuration. This is because $D^{\mathcal{N}}_{j_{_{\mathcal{N}}},j_{_{\mathcal{N}}}}(E^\mathcal{N}_0)=\mathcal{A}^{\mathcal{N}}(E^\mathcal{N}_0)=1$, which is just the $\mathcal{N}$-fermion spectral function evaluated at the ground state energy.
Thus from Eq.~\ref{e3} we see that the $A^{(2-hole)}_{IJ;IJ}(\omega)=\sum_{j_{_{\mathcal{N}}},j^\prime_{_{\mathcal{N}}}
}\mathcal{D}^{\mathcal{N}-2}_{j_{_
{\mathcal{N}}}(IJ)^-;j^\prime_{_{\mathcal{N}}}(IJ)^-}(\omega)$. To ascertain that the peak at $-2\Delta-U=-24$ in Fig.~\ref{f-3} (b) for $U=\Delta=8$  is indeed a two-hole local resonance, we compare the $A^{(2h-TM)}_{L}(\omega)$ with $\mathcal{D}^{\mathcal{N}-2}_{Proj}(\omega)=\sum_{\mathbf{j}_{_
{\mathcal{N}-2}}}\mathcal{D}^{\mathcal{N}-2}_{\mathbf{j}_{_
{\mathcal{N}-2}};\mathbf{j}_{_{\mathcal{N}-2}}}(\omega)$, where the $\mathbf{j}_{_{\mathcal{N}-2}}$ label runs only over the basis states which have the two holes together on any TM site-pair for $U=\Delta=8$. Fig.~\ref{f-3} (c) shows the projected spectral function. We immediately see that the peak of this quantity coincides with that of $A^{(2h-TM)}_{L}(\omega)$. Since $A^{(2h-TM)}_{L}(\omega)$ contains the TM site-pair projected contribution only, it clearly shows that the peak at $-2\Delta-U$ is indeed the local two-hole resonance. Moreover, the trace over the projected basis also shows that the effective bandwidth of the two holes delocalizing over only the TM pairs only is negligible. The effective Hamiltonian approach, as was discussed for the NN interaction model in the previous sub-section, also provides an infinitesimally small bandwidth in the present case.  

\begin{figure*}
\centering{
\includegraphics[width=16cm, height=8cm, clip=true]{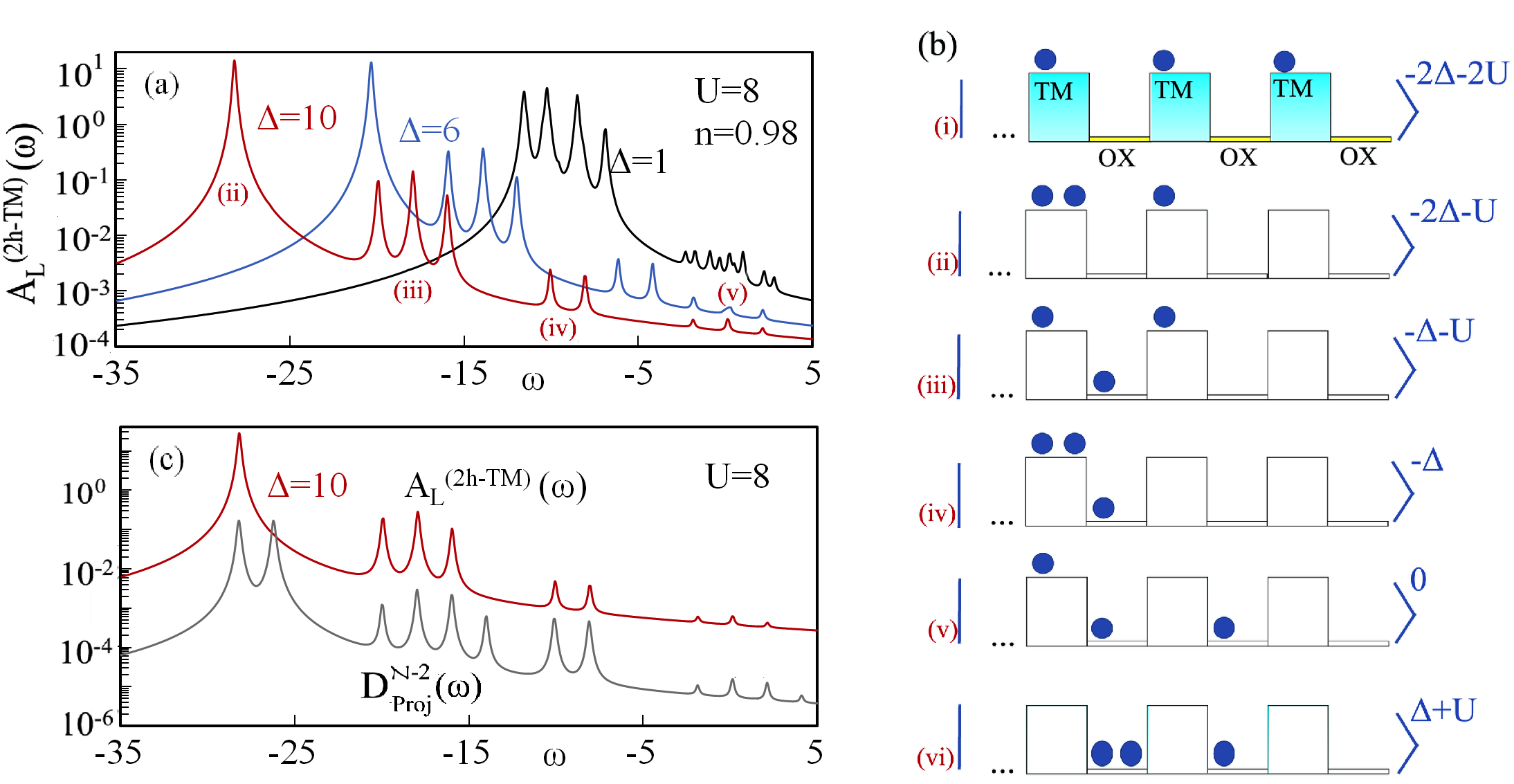}} 
\caption{\textbf{\textit{Local} two-hole spectral function in the almost filled ground state or at small hole doping.} (a) show the local two-hole spectral function for pair of holes created on a TM site-pair. The results are for doping values $n=0.98$, $U=8$, and three $\Delta$ values. In (b), we show the six possible three-hole configurations in increasing order of their potential energies (from top to bottom). The TM  site-pairs are shown as a single level, elevated by $\Delta$ from the OX site-pairs, as labeled in the top configuration. Filled circles denote the holes in an otherwise filled system. The potential energies for each configuration are indicated as superscripts for every state. The schematic only shows representative configurations in every case.   (c) Compares the two-hole spectra (red) and projected three-hole spectral function contribution in grey. The projection involved in the grey curve is discussed in the text.}
\vspace{-0.0cm}
\label{f-4}
\end{figure*}

The almost vanishing bandwidth (within numerical accuracy) leads to the strong spatial localization of $A^{(2h-TM)}_{L}(\omega)$ at the TM site-pair where the two holes were created and stabilizes the '\textit{local} 2-hole resonance' L2HR. This preference of the \textit{local} two-hole resonance also suppresses the other possible feature centered around $-2\Delta-2U$, corresponding to both holes occupying different TM sites, to negligibly small values. 

Reducing $\Delta$ to 4, we see a clear shift of spectral weight from the two-hole resonance primarily to the central feature. The three features still are centered around $-2\Delta-U$, $-\Delta-U$, and zero. However, we notice that the relative height of the central feature also grows roughly by order of magnitude. As for $\Delta=8$, the $\mathcal{D}^{\mathcal{N}-2}_{Proj}(\omega)$ projected onto basis states of two holes on a TM site pair still has the dominant contribution at the peak at $-2\Delta-U$ allowing the interpretation of this feature as the L2HR.

Finally, for $\Delta=0$, the resonance merges with the central feature (centered at $-U$), and the L2HR is lost. On the other hand, if $U=0$ and $\Delta$ is non-zero, the potential energies of the basis states that contribute to $-2\Delta-U$ and $-2\Delta-2U$ become degenerate, and the ground state has comparable contributions from the two holes of TM site pairs as well as then they are singly occupying different TM site-pairs. Hence we conclude that a combination of $U$ and $\Delta$ is needed for stabilizing a well-separated resonance, primarily localized on the TM site-pair in the undoped case. 
\textcolor{black}{For completeness, we briefly discuss the origin of the sub-structures centered around $\omega=-16$ and $\omega=0$ seen in Fig 2 (b) for $U=8$ and $\Delta=8$. The basis state potential energy for one TM hole and one OX hole is $-\Delta-U=-16$ and is zero for both holes on OX. Since the states with two holes on TM hybridize with the above two category of states, they show features around the locations $\omega=-16$ and $\omega=0$. The support of the features on the $\omega$-axis is controlled by the spread of the spectrum belonging to the states of the two categories. Of course, the magnitudes are highly suppressed because large $\Delta$ and $U$. The red (blue) ellipses on the $\omega$-axis in Fig 2 (b) show the eigenvalues of the Hamiltonian diagonalized in the restricted basis containing one TM and one OX holes (two OX holes). We find that these locations agree with the sub-structures around $\omega=-16$ and $\omega=0$. We note that due to large $U$ and $\Delta$ the effect of hybridization which would shift the locations of these sub-structures (usual level repulsion effects) is small. They are likely to be more relevant at smaller $U $ and $\Delta$ values.}

\textit{\underline{(ii) Local two-hole spectrum for small doping:}} In panel Fig~\ref{f-4} (a), we show $A^{(2h-TM)}_{L}(\omega)$ for a ground state of $\mathcal{N}=47$ fermions in $\mathcal{L}=48$ site lattice. The resulting filling, $n\sim 0.98$, is the smallest nontrivial doping possible at this system size, consisting of a single hole in the $\mathcal{N}$-fermion ground state. For the partially filled case, we first work out $Tr\{\mathcal{D}^{\mathcal{N}}_{j_{_{\mathcal{N}}},j_{_{\mathcal{N}}}}(\omega)\}$ and locate the lowest energy peak, which gives the $\mathcal{N}$-fermion ground state energy for the given Hamiltonian. As before, we shift $A_L^{(2h-TM)}(\omega)$ by the ground state energy. We see in panel (a) that the two-hole spectrum develops more features than the filled case due to additional charge fluctuations due to one hole already present in the ground state. 
To analyze $A_L^{(2h-TM)}(\omega)$ shown in panel (a), we first list all the possible three-hole configurations once we create two extra holes. Panel (b) shows the six possible kinds of three-hole configurations in order of increasing energy from (i) to (vi). Superscripts for each configuration denote their potential energies as measured from the lowest possible potential energy of the basis states. The `lowest possible potential energy' corresponds to the potential energy of  $\mathcal{N}$-fermion basis state with the ground state hole on any TM site-pair. 
All basis states belong to one of the six groups; depending on the total occupation of the TM and OX sites, the schematic shows some representative conjugations of each group.
 
For $\Delta=10$, we see four dominant structures centered around $\omega=-2\Delta-U(=-28)$, $\omega=-\Delta-U(=-18)$, $\omega=-\Delta(=-10)$ and $\omega=0$. These correspond to the configurations (ii)-(v) in panel (b). The lowest energy configuration shown in (b) is $-2\Delta-2U$. This feature is strongly penalized, as discussed for the undoped case. 

Further, in panel (a), we find that with reducing $\Delta$, the initially suppressed features gain in magnitude, the centroid of the feature (ii) rapidly approaches (iii), and for $\Delta\leq 2$, it merges with the other structures.
For example, for $\Delta=6$, the features (ii), (iii), and (iv)  move close to each other. There is also an overall shift of all the low-energy features toward zero. 
Finally, for $\Delta=1$, we see that the features (ii), (iii), and (iv) merge.

We now analyze  $A^{(2h-TM)}_{L}(\omega)$ the $n=0.98$ for $\Delta=10$ based on the many-fermion density of states. From Eq.~\ref{e3}, we crucially observe that the $\omega$ dependence of  $A^{(2h-TM)}_{L}(\omega)$ comes \textit{only from $\mathcal{N}-2$ fermion DO}S. In Fig~\ref{f-4} (c), we show $\mathcal{D}^{\mathcal{N}-2}_{Proj}(\omega)=\sum_{\mathbf{j}_{_{\mathcal{N}-2}}}\mathcal{D}^{\mathcal{N}-2}_{\mathbf{j}_{_
{\mathcal{N}-2}};\mathbf{j}_{_{\mathcal{N}-2}}}(\omega)$ in grey, where, the 
$\mathbf{j}_{_{\mathcal{N}-2}}$ label now runs only over the basis states where two holes are on any pair of TM sites and the third hole is on any other TM pair. We choose to project into these basis states since the $\mathcal{N}$-fermion ground state is dominantly made up of the basis states where the ground state hole occupies TM rather than OX. 

To compare, we replot in (c) (red curve) the corresponding $A^{(2h-TM)}_{L}(\omega)$ for $\Delta=10$, from panel (a). Unlike in the undoped case where $D^{\mathcal{N}}_{j_{_{\mathcal{N}}},j_{_{\mathcal{N}}}}(E^\mathcal{N}_0)$ was 1, here we see that even in the presence of a single hole in the ground state, the $D^{\mathcal{N}}_{j_{_{\mathcal{N}}},j^\prime_{_{\mathcal{N}}}}(E^\mathcal{N}_0)$ has a significant effect. In particular, only the low energy peak of  $\mathcal{D}^{\mathcal{N}-2}_{Proj}(\omega)$ at $(\omega=-2\Delta-U=-28)$ corresponding to configuration (ii) of the panel (b) retains its contribution in $A^{(2h-TM)}_{L}(\omega)$. Similar reduced contributions are seen for all other features as well.
This suppression of the features results from the summation over  $j_\mathcal{N}$ and  $j_\mathcal{N}^\prime$ and the fact that the off-diagonal elements of $\mathcal{N}$ and $\mathcal{N}-2$ particle spectral function matrices are not positive definite. We have also explicitly performed restricted summation over the $j_\mathcal{N}$ and  $j_\mathcal{N}^\prime$ in Eq. 1, such that only one of the six possible $(\mathcal{N}-2)$-fermion basis state groups, contribute at a time to $\mathcal{D}^{\mathcal{N}-2}(\omega)$. From this analysis, we have ascertained that apart from the basis states where two holes occupy a single  TM site-pair and the third hole is on any other TM pair, no other basis state has an appreciable contribution to the two-hole spectrum at $\omega=-2\Delta-U$.
A similar analysis for other basis state groups shows highly suppressed contributions to the peak at $-2\Delta-U$ for $\Delta=10$ and $\Delta=6$. For $\Delta=1$, many of the six basis state groups contribute at all energies, wiping out the signature of the  L2HR from the local two-hole spectral function. 

\begin{figure}[!t]
\centering{
\includegraphics[width=8.5cm, height=6.8cm, clip=true]{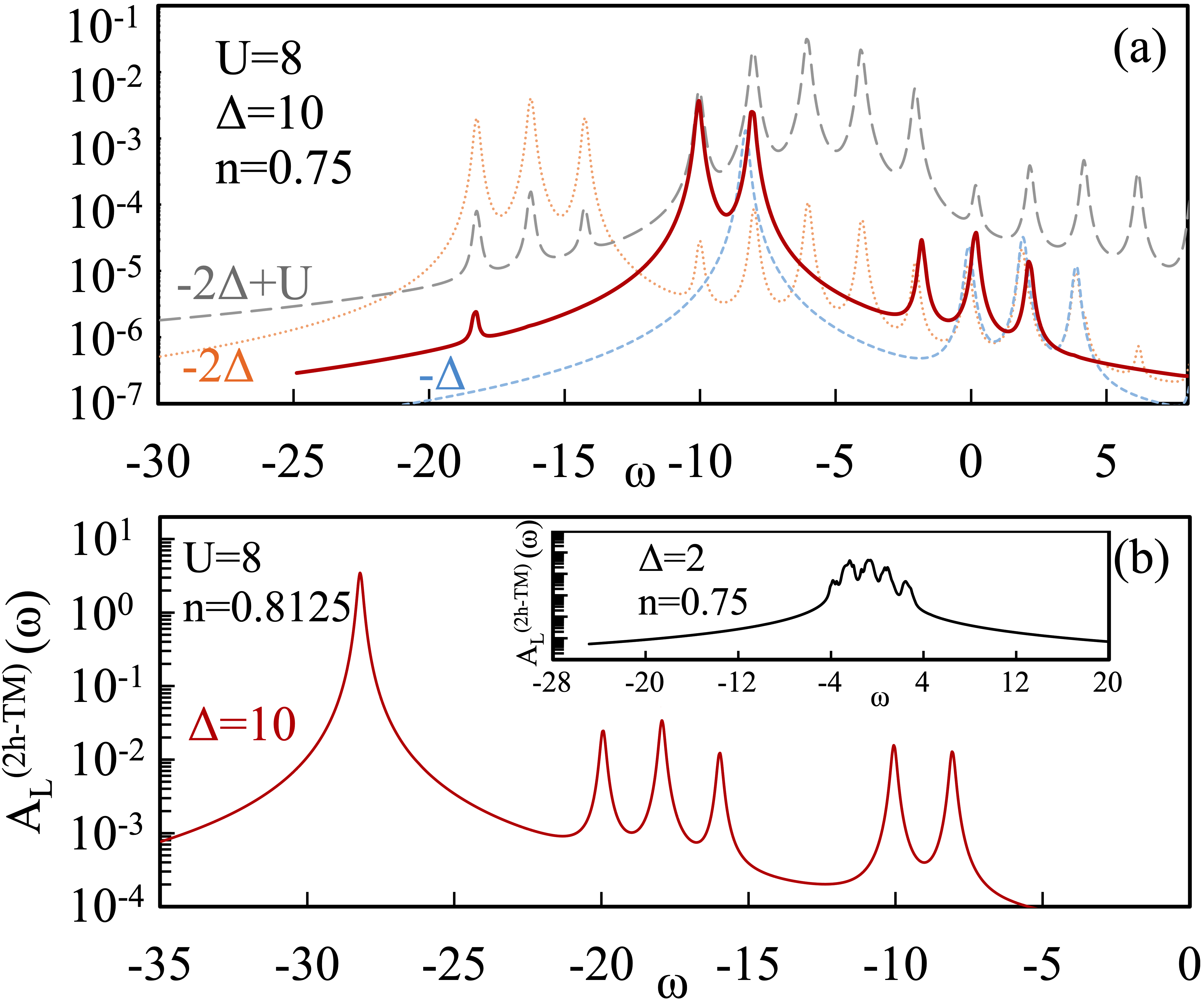}} 
\caption{\textbf{\textit{Local} two-hole spectral function in large hole-doped ground states.} In (a), we show the local two-hole spectral function for pair of holes created on adjacent TM sites for $n=0.75$, for $\Delta=10$ and $U=8$ by the solid line. The dotted (orange), dashed (grey), and fine-dashed (blue) lines show  $(\mathcal{N}-2)$-fermion spectral function projected on basis states with potential energies $-2\Delta$, $-2\Delta+U$ and $-\Delta$ respectively. The main panel in (b) shows the two-hole spectral function for $n=0.8$, which has four holes in the ground state compared to five at $n=0.75$ for $\mathcal{L}=20$, the size studied here. Inset in (b) shows the two-hole spectral function for $n=0.75$ and small $\Delta(=2)$. }
\vspace{-0.0cm}
\label{f-5}
\end{figure}

\underline{\textit{3. Larger hole doping case:}} We now consider $n=0.75$ or 25\% hole doping in the ground state. For reducing computational cost, we limit to $\mathcal{L}=20$ and $\mathcal{N}=15$ or five holes in the ground state. In Fig.~\ref{f-5} (a) we show $A^{(2h-TM)}_{L}(\omega)$ for $\Delta=10$ for  and  $U(=8)$ by the solid red curve.

We also show the $(\mathcal{N}-2)$-fermion spectral function projected onto three sets of $(\mathcal{N}-2)$-fermion basis states with the lowest three potential energies. These three curves depict $\mathcal{D}^{\mathcal{N}-2}_{Proj}(\omega)$, projected onto basis states with potential energies $-2\Delta$, $-2\Delta+U$ and $-\Delta$, measured from the lowest possible basis state potential energy. These correspond to basis-state groups with only two, three, and one TM site-pair doubly occupied by holes, respectively.
Comparing the peak locations of $A^{(2h-TM)}_{L}(\omega)$ with the three  $(\mathcal{N}-2)$-fermion projected spectral function we find that the lowest peak in $A^{(2h-TM)}_{L}(\omega)$ has contributions from basis states with potential energies $-2\Delta$ followed by those with potential energy $-2\Delta+U$. From above, these basis states have two and three TM site-pairs doubly occupied by holes. The most prominent peak in $A^{(2h-TM)}_{L}(\omega)$ has a contribution from basis states with three TM site-pairs doubly occupied by holes and from basis states with two TM site-pairs doubly occupied by holes. In fact, the $(\mathcal{N}-2)$-fermion basis states that have only one TM-pair doubly occupied by holes and well separated from other features are those with potential energy $-\Delta$ (blue fine-dashes line) and contributes to the third features of $A^{(2h-TM)}_{L}(\omega)$. We have explicitly checked that among the 17 possible $(\mathcal{N}-2)$-fermion basis state groups with distinct potential energies, no other groups contribute to the low energy peaks of $A^{(2h-TM)}_{L}(\omega)$.
Thus, the lowest energy peak can no longer be interpreted as L2HR. In contrast in panel (b) we show  $A^{(2h-TM)}_{L}(\omega)$  for $n=0.8$ for the same system size, $\Delta$ and $U$. This case has only one less hole in the $\mathcal{N}$ -fermion ground state or four holes compared to five for $n=0.75$. We see that  $A^{(2h-TM)}_{L}(\omega)$ for $n=0.8$, has a clear low energy feature, whose composition is similar to the low doping cases discussed earlier and can be interpreted as L2HR. We do not repeat the analysis here. This rather drastic spectral weight redistribution with a small change in hole doping can be rationalized in the following manner. 


The two-hole addition on a TM site-pair is only possible on basis states with at least one TM site-pair doubly occupied by two fermions. At $n\leq0.75$, the ground state is dominantly composed of basis states where all fermions reside on OX at large $\Delta$ and $U$. The basis states where two-hole addition only has finite amplitude have to have at least one doubly occupied TM site-pair. However, these  $\mathcal{N}$-fermion basis states have higher potential energies compared to those where the TM is singly occupied or unoccupied and consequently have small contributions to the ground state wave function. For completeness, in the inset of (b), we show $A^{(2h-TM)}_{L}(\omega)$ for small $\Delta(=2)$, which, as expected, does not have any clear resonance peak. In summary, at $n=0.75$, the spectral weight of the two-hole spectral function is strongly distributed among features constructed out of basis states where holes doubly occupy multiple TM site-pairs. We note that these conclusions hold for larger system sizes as well. Regardless of the system size for $n\leq0.75$, there are basis states where no TM site-pairs are doubly occupied with fermions. Since the ground state for large $U$ and $\Delta$ is dominantly constructed out of such basis states, the matrix element for creating two holes on a TM site-pair is significantly suppressed. 


\begin{figure}[!t]
\centering{
\includegraphics[width=8.5cm, height=4.6cm, clip=true]{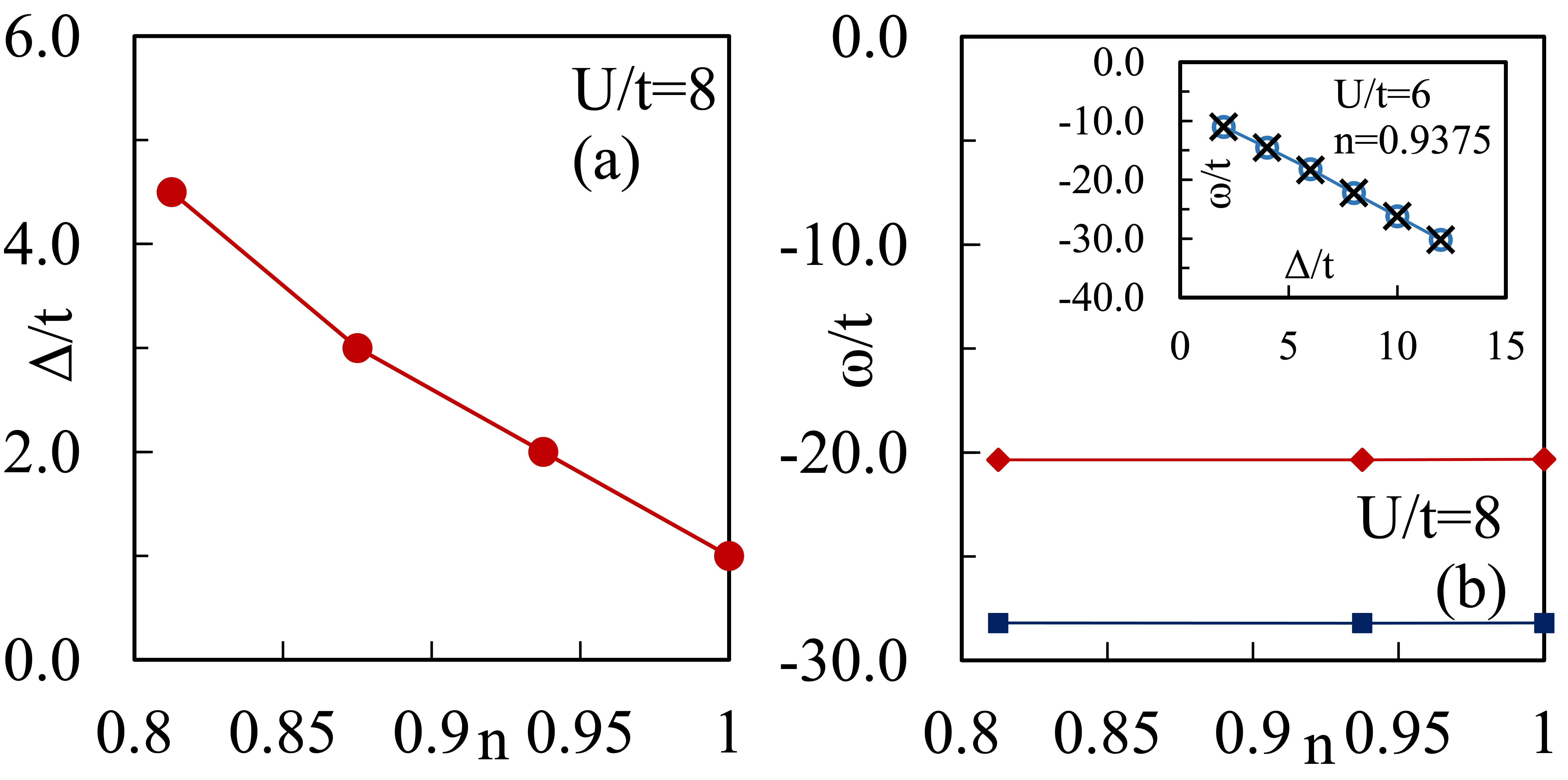}} 
\caption{\textbf{Doping dependence of $\Delta_{crit} $ for local two-hole resonance.} (a) $\Delta-$filling (n) plot showing the critical charge transfer energy for well-defined local two-hole resonance for large $U(=8t)$. (b) The scaling of the energy location of the L2HR feature with $\Delta$ for $\Delta>\Delta_{crit}$. The main plot shows the hole-doping dependence of the L2HR peak location (centroid) for $\Delta=6t$ (diamonds) and 8$t$ (squares) for fixed $U(=8t)$. The inset shows the L2HR centroid locations with varying $\Delta>\Delta_{crit}$  for $U=6t$ at $n=0.9375$ with open symbols. The crosses are obtained from Lanczos.}
\vspace{-0.0cm}
\label{f-6}
\end{figure}
\underline{\textit{Critical $\Delta$ for local two-hole resonance:}} Fig.~\ref{f-6} (a) shows the plot of the critical $\Delta$ needed for a well-defined L2HR as a function of hole doping of the ground state. \textcolor{black}{The critical $\Delta$ is defined to the $\Delta$ value for which a single low-energy resonance (centered at $-2\Delta-U$) is pulled out of the rest of the two-hole spectral function features.}
The $U$ is set to be larger than the kinetic energy bandwidth, as is typical for  3d TM oxides. 
We also consider positive $\Delta$, which  apart from small bandwidth dependent corrections, dictates that holes doped to create partially filled ground-state  prefer to occupy the TM site-pairs, similar to doped Mott insulators. In the regime of $\Delta$ where we have a stable L2HR, we numerically observe this to be true.
Also, a similar $\Delta<0$ analysis can be carried out, mimicking a negative-charge transfer situation. However, in partially filled bands, if added holes prefer OX, the matrix element for creating two holes on TM is very small. We thus only consider positive $\Delta$. At zero hole doping, the $\Delta_{crit}/t\sim 1$ and increases linearly with hole doping, up to about slightly less than 75\% band filling. Beyond that, as discussed for $n\leq 0.75$ above,  there is a drastic redistribution of spectral weight in the local two-hole spectral function destabilizing the  L2HR for any $\Delta$. While for $n>0.75$, two-hole resonance can be stabilized for $\Delta>\Delta_{crit}$.

\underline{\textit{Relevance for L2HR in AES:}}
As the introduction mentions, the Core-Valence-Valence AES, a core hole, decays into two final state holes in the valence band and an Auger electron.
\textcolor{black}{ If the interaction between two final state holes is weak, the holes delocalize over the lattice screened from each other, with the AES or the two-hole spectral function closely resembling the convolution of two holes. }
In this case, the AES signal is proportional to the two-hole convolution. In the strongly correlated limit, the strong interaction can localize the electrons at the atomic site where they were created, shifting the energy higher than the screened two-hole energy by the Coulomb correlation. This fact is central to the Cini-Sawatzky theory for analyzing the AES experiments. They calculated the two-hole local Green's function for holes added to a filled $d$ band exactly in terms of the one-hole Green's function. The approach shows that a split-off resonance with $d^8$ multiplet structure is generated whose energy location is $U_{dd}$, from the energy where two holes are on different sites or the band part of the spectrum. This theory captures the experimental spectra for filled 3$d$ band systems such as Cu and Zn quite well. However, the major drawback of the theory is that the exact result holds when the one-hole Green's functions are known. For example, even for Ni, which is $3d^8$, the one-hole Green's function added in a partially filled 3$d$ band is non-trivial and direct input for XPS is needed\cite{Wandelt_Mondio_Cubiotti_1989}.  

As in the Cini-Sawatzky theory, impurity calculations \cite{dd-auger} in the filled limit shows a satellite peak separated from the band-like features by the local correlation strength. In this limit, the impurity results also show a simple linear relation of the satellite's location with $U$. However, adding small-hole doping before the Auger two-hole insertion strongly distorts the picture even when core hole effects are not considered. The linear dependence of the satellite peak is lost, and the satellite shifts towards the band-like part with increasing $U$. Finally, the effects of $\Delta$ have not been explored systematically at partial filling within the impurity studies to the best of our knowledge. 

From Fig.~\ref{f-6}  (a), we already see that the L2HR, or the AES satellite, is stable beyond a filling dependent $\Delta_{crit}$. We briefly comment on the $\Delta$ dependence of the energy location of L2HR  at large $U$. In the inset in  Fig.~\ref{f-6}  (b), we see that for hole doping less than 25\%,   ($n>0.75$), the L2HR location exhibits a simple linear dependence of $-U-2\Delta$  for large fixed $U(=6)$ for $n<1$. This is representative of the behavior for all $n>0.75$ and $\Delta$ greater than the n-dependent critical value. This is analogous to the linear dependence on $U$ in the literature previously reported in the undoped case \cite{dd-auger}. However, unlike the impurity approach, retaining the full lattice symmetry extends the regime where the L2HR has a linear dependence on $U$ above 75\% band-filling. 

Since $\Delta$ can be determined from X-ray techniques such as X-ray photo-electron spectroscopy (XPS)\cite{xps}, $U$ can be ascertained for the location of the L2HR. The main panel Fig.~\ref{f-6}  (b) shows the \textit{filling-independence} of the energy location of the L2HR for $0.75<n\leq1$. Representative data are shown for  $U=6$ and $U=8$. We finally note that all results presented here using the low-memory approach are bench-marked against Lanczos-based diagonalization. As an example, in the inset in  Fig.~\ref{f-6}  (b), we show that the Lanczos-based L2HR peak (crosses) are in excellent agreement with our method.

\textcolor{black}{We conclude this section with general remarks summarizing the interplay of $\Delta$ and $U$, expectations in two dimensions, and spin-full cases. For two spinless holes added in a filled ground state of our one-dimensional model, we first consider the non-interacting problem, including non-zero $\Delta$. When  $U$ crosses the minimum magnitude of the non-interacting two-hole band, a spectral feature is split from the non-interacting continuum. However, the split-off feature contains contributions from states with two holes on a single TM site-pair, one hole each on TM and OX, both holes on OX, and both individually occupying two different site-pairs. These features are  centered around $-2\Delta-U$ and $-\Delta-U$, 0, and  $-2\Delta-2U$, as discussed earlier. Thus, to separate the L2HR centered around $-2\Delta-U$ from the other features, $\Delta$ has to be suitably chosen. The separation between these features depends on the feature centroids and the bandwidth of individual features. Thus, we have employed large $U$ in our paper so that $\Delta$ can be chosen to isolate the L2HR.} \\
\textcolor{black}{With this understanding, we can consider the two-dimensional extension of our spinless fermion model with $2\times2$ plaquettes of TM and OX. Since two-hole bandwidth is larger in two dimensions, the critical $U$ (for a fixed $\Delta)$ to create split-off features is expected to be larger than in one dimension. Similarly, since the individual features will have more delocalization paths in two dimensions, their bandwidths are expected to be larger. Thus, for the same fixed large $U$ in one and two dimensions, the $\Delta_{crit}$ needed to separate the L2HR is expected to be greater in two dimensions. The same trend should also hold for two-hole added in partially filled ground states. Moreover, due to the  2$\times$2 plaquette structure, the two-hole spectral function is expected to have a sub-structure even if two holes are spatially localized on a single plaquette. Thus, the occupation of a TM plaquette by two holes will reduce the energy from the filled ground state by $-4U-2\Delta$ for holes occupying diagonal sites of the plaquette, followed by $-3U-2\Delta$ when the two holes occupy adjacent sites on a TM plaquette. For one hole on OX, the energy reduction is $-2U-\Delta$ and 0 for both holes on OX. A careful analysis would be needed to determine the critical $\Delta$  at large $U$ beyond the above general expectations, for the feature centroids.  \\
The most straightforward material realistic extension of the model can be done by considering the well-known ionic Hubbard model \cite{ihm-1,ihm-2} with spin-full fermion in two dimensions. The model consists of TM and OX sites (to use the language in the paper) and can be studied in the case with onsite-Hubbard $U$ and $\Delta$ on TM sites (which can be occupied by two electrons of opposite spins) and non-interacting OX sites with zero onsite potential. The model is of relevance to strongly correlated double perovskites\cite{ihm-3}. The $U_{crit}$ for two holes added in a filled band will be approximately $U\geq\sqrt{\Delta^{2}+64t^2}$, twice the non-interacting one-hole bandwidth. For $U$ set to be much larger than this critical value will allow adjusting $\Delta$ to separate an onsite TM local two-hole resonance.  \\}

\section{Conclusions}

\textcolor{black}{We have investigated the impact of strong correlation and charge transfer effects on the stability of local two-hole spectral function in partially filled ground states using exact numerical techniques retaining full lattice symmetry.} We have established that in contrast to only NN repulsive interaction, in the model with TM and OX motif, both interaction and charge transfer energy is necessary for stabilizing a two-hole resonance with vanishing small bandwidth even in a filled ground state. For strong correlations, we have uncovered a charge transfer regime where the L2HR is stable for a wide regime of partially filled ground states, from filled to about 75\% filling. At 75\%band filling, the local two-hole spectral function shows an abrupt and dramatic spectral weight redistribution, destabilizing the signature of the L2HR in the local two-hole spectral function. For $n\leq0.75$, the L2HR cannot be recovered for any $U$ and $\Delta$, unlike for $n>0.75$. Finally, we have shown that the location of the L2HR has a linear relation of both $\Delta$ and $U$, for $\Delta>\Delta_{crit}$ 
In this regime, one could use the energy location of the L2HR to extract the value of TM interaction strength, which is an extension of the Cini-Sawatzky type approach for ascertaining correlation strength from L2HR  to partial band filling. Further, since our result produces the exact two-hole spectral function, the approach provides a way to fit the relevant experimental data in any parameter regime. 
Our numerical scheme can also handle the inclusion of core holes and can handle local multiplet structures and, in the future, will allow us to make a realistic comparison with AES experiments.

\section{Acknowledgements:}
All computations were preformed in the NOETHER, VIRGO and KALINGA high performance clusters at NISER. We acknowledge funding from the Department of Atomic Energy, India under Project No. 12-R\&D-NIS-5.00-0100. A.M. would also like to acknowledge  SERB-MATRICS grant (Grant No. MTR/2022/000636)  from the Science and Engineering Research Board (SERB) for funding. 
\appendix*
 \setcounter{equation}{0}  
\begin{center}
	\textbf{APPENDIX}
\end{center}

\subsection{Many-fermion formalism}

For two holes created on a pair of lattice sites $(I,J)$ in a $\mathcal{N}$-fermion ground-state $\ket{\psi_0^\mathcal{N}}$ and subsequently destroyed at a later time from the same site-pair, the two-hole retarded Green's function, in the frequency domain is given by:
\begin{eqnarray}
&G^{{(2h)}}_{IJ;IJ}&(\omega)=\sum_{j_\mathcal{N},j_\mathcal{N}^\prime}\langle{\psi_0^\mathcal{N}}\ket{j_\mathcal{N}}\langle{j_\mathcal{N}^\prime}\ket{\psi_0^\mathcal{N}}\nonumber\\ 
&\times&\bra{j_\mathcal{N}} c^\dagger_{I}c^\dagger_{J} ((\omega+ i\eta)I-H)^{-1}c_{I}c_{J}\ket{j_\mathcal{N}^\prime}]
\label{se1}
\end{eqnarray}
In the above we have inserted a complete set of $\mathcal{N}$-fermion real-space basis sets $\{|j_\mathcal{N}\rangle\}$ and $\{|j^{\prime}_\mathcal{N}\rangle\}$. The two-hole spectral function is obtained from the imaginary part of the above expression. We first provide a way to obtain the pre-factors $\bra{\psi_0^\mathcal{N}}\ket{j_\mathcal{N}}\bra{j_\mathcal{N}^\prime}\ket{\psi_0^\mathcal{N}}$ extract the factors in Eq.~\ref{se1} from the $\mathcal{N}$-fermion Green's function introduced in the Methods section in the paper. 
For this, we note that from the imaginary part of $\mathcal{N}$-fermion, Green's function in the Lehmann representation can be expressed as: 

\begin{eqnarray}
-\frac{1}{\pi}\Im{\mathcal{G^{\mathcal{N}}}_{j_{\mathcal{N}};j^\prime_{\mathcal{N}}}(\omega)}=\sum_{\lambda^{\mathcal{N}}}\bra{j_{\mathcal{N}}}\ket{\lambda^{\mathcal{N}}}\bra{\lambda^{\mathcal{N}}}\ket{j^\prime_{\mathcal{N}}}\nonumber\\ \times\delta(\omega-E^{\mathcal{N}}_{\lambda}) ~~~~~~~~~~~~~~~
\label{se3}
\end{eqnarray}
which implies that, 
\begin{eqnarray}
-\frac{1}{\pi}\Im{\mathcal{G^{\mathcal{N}}}_{j_{\mathcal{N}};j^\prime_{\mathcal{N}}}(\omega=E^\mathcal{N}_0)}=\bra{j_{\mathcal{N}}}\ket{\psi_0^\mathcal{N}}\bra{\psi_0^\mathcal{N}}\ket{j^\prime_{\mathcal{N}}}\nonumber\\=D^{\mathcal{N}}_{j_{\mathcal{N}};j^\prime_{\mathcal{N}}}(E^\mathcal{N}_0)~~~~~~~~~~~~~~~~~~~~~~~~~~~~~~~~~~~~~~~~~~~~~~~~
\label{se4}
\end{eqnarray}
in the $\eta\rightarrow 0$ limit, where $D^{\mathcal{N}}_{j_{\mathcal{N}};j^\prime_{\mathcal{N}}}(\omega)$ is the $\mathcal{N}-$fermion spectral function matrix. We note that for the above holds for $\mathcal{G}^{\mathcal{N}}_{j_{\mathcal{N}};j_{\mathcal{N}}^\prime}(\omega)=\mathcal{G}^{\mathcal{N}}_{j_{\mathcal{N}}^\prime;j_{\mathcal{N}}}(\omega)$ which is true for equilibrium many-fermion problems. The remaining part of Eq.\ref {se1} are matrix elements of $(\mathcal{N}-2)$-fermion spectral function matrix $[\mathcal{D}^{\mathcal{N}-2}(\omega)]$. Thus, the two hole real-space spectral function, 
$A^{(2h)}_{IJ;IJ}(\omega)\equiv -1/\pi Im\{G^{{(2h)}}_{IJ;IJ}(\omega)\}$ can be expressed as: 
\begin{align}
A^{(2h)}_{IJ;IJ}(\omega)=\sum_{j_{_{\mathcal{N}}},j^\prime_{_{\mathcal{N}}}
}\mathcal{D}^{\mathcal{N}}_{j_{_{\mathcal{N}}},j^\prime_{_{\mathcal{N}}}}(E^
{\mathcal{N}}_0)\mathcal{D}^{\mathcal{N}-2}_{j_{_
{\mathcal{N}}}(IJ)^-;j^\prime_{_{\mathcal{N}}}(IJ)^-}(\omega)
\label{e6}
\end{align}
While this form is true for any site-pair $(I,J)$, we particularly focus of the case when $I$ and $J$ are NN of each other. In this case, we refer to the  two-holes spectral function as 
 \textit{local} two-holes spectral function $A^{2-hole}_{IJ;IJ}(\omega)\equiv A_L^{2-hole}(\omega)$ .
From Eq.~\ref{e6} we first notice that the calculation of real space two-hole  spectral function $A^{(2h)}_L(\omega)$ involves elements of the $\mathcal{N}$-fermion spectral function matrix, $\mathcal{D}^{\mathcal{N}}
_{j_{_{\mathcal{N}}},j^\prime_{_{\mathcal{N}}}}(\omega)$ evaluated at $\omega=E^{\mathcal{N}}_0$ or at the many-fermion ground-state energy. Different elements of $\mathcal{D}^{\mathcal{N}}_{j_{_{\mathcal{N}}},j^\prime_{_{\mathcal{N}}}}(\omega)$ are extracted from the many fermion Green's function $\mathcal{G}^{R(\mathcal{N})}_{j_{_{\mathcal{N}}};j^\prime_{_{\mathcal{N}}}}(\omega)=\langle j_{_{\mathcal{N}}}|\mathcal{\hat{G}(\omega)} |j^\prime_{_{\mathcal{N}}}\rangle$, evaluated between the $\mathcal{N}$-fermion basis elements, $(|j_{_{\mathcal{N}}}\rangle)^\dagger$ and $|j_{_{\mathcal{N}}}^\prime\rangle$ at $\omega=E^{\mathcal{N}}_0$. 
The $(\mathcal{N}-2)$-fermion spectral function matrix elements required in Eq.~\ref{e6}, are similarly extracted from $[\mathcal{G}^{\mathcal{N}-2}(\omega)]$.
The $j_{_{\mathcal{N}}},j^\prime_{_{\mathcal{N}}}$ indices run over all the $\mathcal{N}$-fermion basis-states, while the relation  $|j^\prime_{_{\mathcal{N}}}(IJ)^-\rangle\equiv c_Ic_J|j_{_{\mathcal{N}}}^\prime\rangle$ defines the $(\mathcal{N}-2)$-fermion basis indices in Eq.~\ref{e6}. The unprimed indices refer to corresponding conjugate states and are defined analogously. 

Finally, we note that two-particle excitations in partially-filled bands can be computed from $\mathcal{N}$ and  $(\mathcal{N}+2)$-fermion spectral function matrices. Similarly, one (particle/hole) photo-emission excitations can be computed from $\mathcal{N}$ and  ($\mathcal{N}+1/\mathcal{N}-1$) fermion spectral function matrices.

\subsection{Fock-space Recursive Green's function scheme} 
Here we briefly discuss the F-RGF scheme \cite{prabhakar2022memory}. We have a $\mathcal{L}$ site chain with periodic boundary conditions containing $\mathcal{N}$ spinless fermions. We divide the lattice into two halves and label all $\mathcal{N}$-fermion states by $|n_l,n_r\rangle$. Here $n_l$ ($n_r$) refers to the number of fermions in the left (right) half. Under nearest neighbor hopping, either ($n_l,n_r$) is conserved, implying no hopping between the two halves or change only by $\pm 1$ if fermions hop between the two halves. Thus, the  $\mathcal{N}$-fermion Hilbert space can be decomposed into a direct sum of $(\mathcal{N}+1)$ `Fock-space sectors' with fixed ($n_l,n_r$) and hopping matrices connecting them. The later connects $|n_l,n_r\rangle$ to $|n_l\pm 1,n_r\mp 1\rangle$. We label the 'Fock-space-sector' with ($n_l,n_r$) occupations by the $\alpha_{n_l+1}$. The Hamiltonian for a particular Fock-space sector $\alpha_i$ is denoted by $H(\alpha_i)$, which contains $n_l=i-1$ and $n_r=\mathcal{N}-n_l$ in the left and right halves. The hopping matrices that connect nearest neighbour sectors $\alpha_i$ and $\alpha_{j}$ sectors is denoted by $[\tau]_{\alpha_i,\alpha_{j}}$. Since the connections are NN, the Hamiltonian has a tridiagonal representation in the Fock-space sector representation. Due to this, the inverse of the resolvent operator $\omega+i\eta-H$ is also tridiagonal.

The main task is to obtain the resolvent operator or the $\mathcal{N}$-fermion Green's function. We will exploit the tridiagonal form of $H$ and the inverse resolvent $\omega+i\eta-H$ for the inversion. The tridiagonal representation with NN Fock-space sector coupling matrices and sector Hamiltonians is similar to a `matrix-valued' one-dimensional lattice. Due to this, it is natural to generalize the well-known Recursive Green's function (RGF) \cite{regf,parallel-rgf,block-inversion} to this one-dimensional lattice in the Fock space. We define a \textit{disconnected} Green's function matrix for a sector $\alpha_i$ by $[\mathcal{G}^{0}]^{-1}_{\alpha_i\alpha_i}(\omega)\equiv\omega-H(\alpha_i)+i\eta$. As mentioned above, $H({\alpha_i})$ represents $H$ in the basis elements belonging only to the sector $\alpha_i$.  
The dimension of $[\mathcal{G}^{0R(\mathcal{N})}]^{-1}_{\alpha_i\alpha_i}(\omega)$ and $H({\alpha_i})$ is 
$ \Perms{i-1}{\mathcal{L}/2}\times \Perms{\mathcal{N}-i+1}{\mathcal{L}/2}$. This is because for $\alpha_i$ sector, we have $n_l=i-1$ and $n_r=\mathcal{N}-n_l$. These matrices contain all interaction and hopping terms connecting the states that preserve the $(n_l,n_r)$-pair. The NN sector connecting matrices, between adjacent ${\alpha_i}$ and ${\alpha_i+1}$ denoted by $[\uptau]_{\alpha_{i},\alpha_{i+1}}$ has the dimension, $(\Perms{\mathcal{N}-i+1}{\mathcal{L}/2}\times \Perms{i}{\mathcal{L}/2})$. 

We also need to introduce an intermediate, \textit{forward connected} Green's function matrix, $[\mathcal{G}^{F}]_{\alpha_i\alpha_i}$, defined in Eq.~\ref{rfg-1}. Below we will show how to obtain retarded Green's function blocks $[\mathcal{G}^{\mathcal{N}}]_{\alpha_i\alpha_i}$,
by a forward and backward recursion involving $[\mathcal{G}^{F}]_{\alpha_i\alpha_i}$, $[\mathcal{G}^{0}]^{-1}_{\alpha_i\alpha_i}$ and $[\uptau]_{\alpha_i,\alpha_j}$. For notational brevity, we have suppressed the $\omega$ arguments. The recursive algorithm applied to the Fock space lattice has the following steps: 

(a) At first, the \textit{forward connected} Green's function is calculated by the following recursive equation:
\begin{eqnarray}
	[\mathcal{G}^{F}]_{\alpha_i\alpha_i}^{-1}=[\mathcal{G}^{0}]^{-1}_{\alpha_i\alpha_i}-[\uptau]_{\alpha_i\alpha_{i-1}}[\mathcal{G}^{F}]_{\alpha_{i-1}\alpha_{i-1}}[\uptau]_{\alpha_{i-1}\alpha_i}\nonumber
\label{rfg-1}
\end{eqnarray}
For a system with periodic or open boundary conditions, we start with the 
$\alpha_1$ sector, with $(n_l=0,n_r=\mathcal{N})$. Since there is no block to its left, from the above equation, $[\mathcal{G}^{F}]_{\alpha_1\alpha_1}=[\mathcal{G}^{0}]_{\alpha_1\alpha_1}$. Starting from this, we obtain all other diagonal blocks of \textit{forward connected} Green's function up to $[\mathcal{G}^{F}]_{\alpha_{\mathcal{N}+1}\alpha_{\mathcal{N}+1}}$, which is the $\alpha_{\mathcal{N}+1}^{th}$ block. Since there are no further blocks, it can be shown that  $[\mathcal{G}^{F}]_{\alpha_{\mathcal{N}+1}\alpha_{\mathcal{N}+1}}= [\mathcal{G}]_{\alpha_{\mathcal{N}+1}\alpha_{\mathcal{N}+1}}$, the retarded Green's function of the $\alpha_{\mathcal{N}+1}$ block \cite{regf}.

(b) From $[\mathcal{G}]_{\alpha_{\mathcal{N}+1}\alpha_{\mathcal{N}+1}}$, all other diagonal blocks of the retarded Green's function can be obtained by a backward recursion equation,
\begin{align}
	[\mathcal{G}]_{\alpha_{i-1}\alpha_{i-1}}=[\mathcal{G}^{F}]_{\alpha_{i-1}\alpha_{i-1}}~~~~~~~~~~~~~~~~~~~~~~~~~~~~~~~~\nonumber\\
	\times(I+[\uptau]_{\alpha_{i-1}\alpha_i}[\mathcal{G}]_{\alpha_{i}\alpha_{i}}[\uptau]_{\alpha_i\alpha_{i-1}}[\mathcal{G}^{F}]_{\alpha_{i-1}\alpha_{i-1}})\nonumber
\end{align}

(c) From the diagonal blocks of the retarded Green's function, we can calculate all off-diagonal blocks by the recursive relation,
\begin{align}
[\mathcal{G}]_{\alpha_i\alpha_j}|_{\alpha_i<\alpha_j}=-[\mathcal{G}^{F}]_{\alpha_i\alpha_i}[\uptau]_{\alpha_i\alpha_{i+1}}[\mathcal{G}]_{\alpha_{i+1}\alpha_j}\nonumber
\end{align}
 
We note that matrix inversions are \textit{only needed in the forward recursion}, and the largest matrix dimension that needs to be inverted is for the $\alpha_{\mathcal{N}/2}$ block. As seen from the above equations, we need two matrices of the dimension the sector $\alpha_i$ and $\alpha_i$ at the $i^{th}$ step of the recursion. Any matrix multiplication to obtain $[\uptau]_{\alpha_i\alpha_{i-1}}[\mathcal{G}^{F}]_{\alpha_{i-1}\alpha_{i-1}}[\uptau]_{\alpha_{i-1}\alpha_i}$ are calculated in a manner that the connection matrices and the relevant Green's function matrices are not allocated in the memory simultaneously. Finally, each frequency point is calculated independently, adding no significant memory overhead.

As detailed in our recent paper \cite{prabhakar2022memory}, due to these features, the requirements of F-RGF, the exponential growth of the Hilbert space, is suppressed by ($1/\mathcal{L}$). This leads to the following major practical advantage. For, say, for $N=20$ and half-filling the current state-of-art, this amounts to a reduction of RAM from 512Gb to a mere 160Gb. It shows that we can perform state-of-the-art calculations at a fractional memory at other fillings as well, allowing access to large system sizes at currently available resources.

\subsection{Bandwidth of two-hole bound-pair}
\textcolor{black}{We estimate the bandwidth of two-hole bound-pair by constructing an effective Hamiltonian by dividing the basis vectors into two groups:} (i) all states with two holes on NN sites and (ii) all other states. We construct the Hamiltonian for these two groups separately, $H_{2h}$ and $H_{r}$, respectively. The full Hamiltonian is a direct sum of these two apart from the connection terms between them. Thus $H$ is given by:
$$
H=\begin{bmatrix} 
H_{2h} & H_{c}  \\
H_{c} & H_{r} \\
\end{bmatrix}
\quad
$$

Here $H_c$ are the hopping terms that connect states of $H_{2h}$ and $H_{r}$.
Using standard manipulations, we can write,
\begin{equation}
H^{eff}_{2h}(\omega)=\omega-[[\omega-H_{2h}]-H_{c} \frac{1}{[\omega-H_{r}]} H_{c}]
\end{equation}
We set $\omega=-3U$ the potential energy of all states with two holes on NN sites to obtain the low-energy effective Hamiltonian. We note that this procedure is valid only if there is a clear energy separation between the states of $H_{2h}$ and $H_{r}$.
The bandwidth in the inset of Fig. 1 (dashed line) in the paper is the bandwidth of the spectrum obtained by diagonalizing $H^{eff}_{2h}(-3U)$.  \textcolor{black}{$D_{BW}$ at large $U$ can also be estimated analytically by standard perturbation approach to order $O(t^2/U)$ and is found to be $4t^2/U$.}

\bibliography{bibliography.bib}
\end{document}